\documentclass[preprint,showpacs,preprintnumbers,showkeys,a4paper,amsfonts,amsmath,aps,jap]{revtex4-1}
\usepackage{dcolumn,amssymb,xspace,multirow}
\usepackage{graphicx}
\usepackage{bm}

\begin{document}

\title{Oxygen vacancy and E$_C~-$ 1 eV electron trap in ZnO}

\author{Gauthier~CHICOT $^{1,2}$}
\author{Pierre~MURET $^{1,2}$}
\author{Julien~PERNOT $^{1,2,3}$}
\email{julien.pernot@neel.cnrs.fr}
\affiliation{$^1$Univ. Grenoble Alpes, Inst NEEL, F-38042 Grenoble, France}
\affiliation{$^2$CNRS, Inst NEEL, F-38042 Grenoble, France}
\affiliation{$^3$Institut Universitaire de France, 103 boulevard Saint-Michel, F-75005 Paris, France}

\author{Jean-Louis SANTAILLER$^{4}$, Guy FEUILLET$^{4}$}
\affiliation{$^{4}$CEA-LETI, Minatec Campus, 17 rue des Martyrs, 38054 Grenoble Cedex 9, France}

\begin{abstract}
Fourier transform deep level transient spectroscopy has been performed between 80 K and 550 K in five $n-$type ZnO samples grown by different techniques. The capture cross section and ionization energy of four electron traps have been deduced from Arrhenius diagrams. A trap 1 eV below the conduction band edge is systematically observed in the five samples with a large apparent capture cross section for electrons ($1.6 \pm 0.4 \times 10^{-13}$ cm$^{2}$) indicating a donor character. The assignment of this deep level to the oxygen vacancy is discussed on the basis of available theoretical predictions.
\end{abstract}

\maketitle

\section{Introduction}
ZnO is a very attractive semiconductor for optoelectronic uses. Its direct wide band gap (3.37 eV) and large binding exciton energy (60 meV) allow ZnO to compete with GaN for light emitting diode (LED) applications in the UV spectrum. To fabricate LED devices, the $n$- and $p-$doping processes must be fully mastered in order to control the conductivity and type of the active layers. Shallow donor levels responsible for the residual $n$-type conductivity of as-grown materials are commonly attributed to native point defects, hydrogen or III elements (like Al, Ga or In) of the periodic table. Formerly, the oxygen vacancy was believed to be one of these shallow states. However, recent theoretical works indicated that the oxygen vacancy (V$_\textrm{O}$) is not a shallow donor level but a deep donor level with a negative U behavior and a (2+/0) charge transition in the energy range 1-2 eV below the conduction band edge $E_C$ \cite{Janotti2005, Erhart, Janotti2007, Oba, Lany1, Lany2}. Based on considerations about the V$_\textrm{O}$ formation energy, some authors found that the concentration of this defect should be low in as-grown $n-$type materials \cite{Janotti2005,Janotti2007}, as confirmed by Electron Paramagnetic Resonance (EPR) experiment in which V$_\textrm{O}$ is detected only after irradiation treatment. The goal of this work is to detect the V$_\textrm{O}$ in $n-$type ZnO crystals grown by different techniques, one of which being implanted, using Fourier transform deep level transient spectroscopy (FT-DLTS) technique performed in a wide temperature range (80 K-550 K). Indeed FT-DLTS is a more sensitive technique than EPR because its sensitivity is at least one part per thousand of the background doping concentration, thus allowing the detection of trap concentrations as low as $10^{13}$ cm$^{-3}$.

This article is organized as follows. In a first part, the experimental details and the FT-DLTS spectra of the five samples grown by different techniques are described. In a second part, the Arrhenius diagrams are analyzed. Finally, the properties of the deepest trap at $E_C-1$ eV are discussed and an assignment to one of the electronic transitions taking place in the oxygen vacancy is shown to be plausible.

\section{Experimental details}
Five (000$\overline{1}$) oriented ZnO samples were investigated in this work : sample $\#1$ is grown by Chemical assisted Vapour phase Transport (CVT) on a ZnO substrate,  sample $\#2$ is a CVT crystal grown on sapphire, samples $\#3$ and $\#4$ are HydroThermal (HT) ZnO crystals and sample $\#5$ is a HT Nitrogen implanted one. The samples $\#1$ and  $\#2$ were grown at 1030$\,^{\circ}$C \cite{Santailler2010} and then annealed at 1100$\,^{\circ}$C during one hour.  The HT samples $\#3$, $\#4$ and  $\#5$ were annealed at 1100$\,^{\circ}$C. Then, the sample $\#5$ was implanted with Nitrogen atoms (multi-implantation with energy ranging between 50 and 200 keV and a total dose of 2.2$\times$10$^{15}$ cm$^{-2}$) and post-annealed at 900$^{\circ}$C. The characteristics of the five samples are summarized in Table \ref{tab:descriptifEch}. All the samples were cleaned with organic solvents before being treated by Remote Oxygen Plasma (ROP). Pt Schottky contacts (50 nm thick and 500 $\mu$m in diameter) were evaporated on the O face of the five samples and full sheet Ti/Au ohmic contacts (20 nm/80 nm) were evaporated on the whole Zn face in order to fabricate the ohmic contact of the diodes. 

Capacitance voltage C(V) and deep level transient spectroscopy measurements were performed with Phystech FT1030 hardware and software. The internal bridge operates at 1 MHz, a measurement frequency which has been checked to be lower than the cut-off frequency of all the diodes. FT-DLTS spectra were obtained from the fast Fourier transform (FFT) of the capacitive transients \cite{weiss_deep_1988}, delivering up to 28 Fourier coefficients for each time window. Current voltage (I(V)) measurements were firstly achieved to check the rectifying behavior of Pt contacts and the leakage current at different temperatures. C(V) measurements were then performed using reverse bias voltage to determine the effective doping level $N_{d}-N_{a}$. Finally, FT-DLTS analysis were performed between 80 K and 550 K using reverse bias  $U_{r}=$ -2 V for sample $\#1$ $\#2$ $\#4$ $\#5$,  $U_{r}=$-4 V for sample $\#3$ and a pulse voltage ($U_{p}$) of 0 V for all the samples. For the five samples, different times windows ($T_{w}$) ranging between 1 ms and 1 s were used in order to collect numerous data, thus improving the accuracy of the Arrhenius diagram. 

\section{Results, analysis and discussion}
\subsection{Fourier transform deep level transient spectroscopy}
\begin{figure}
\centering
\includegraphics[width=0.9\linewidth]{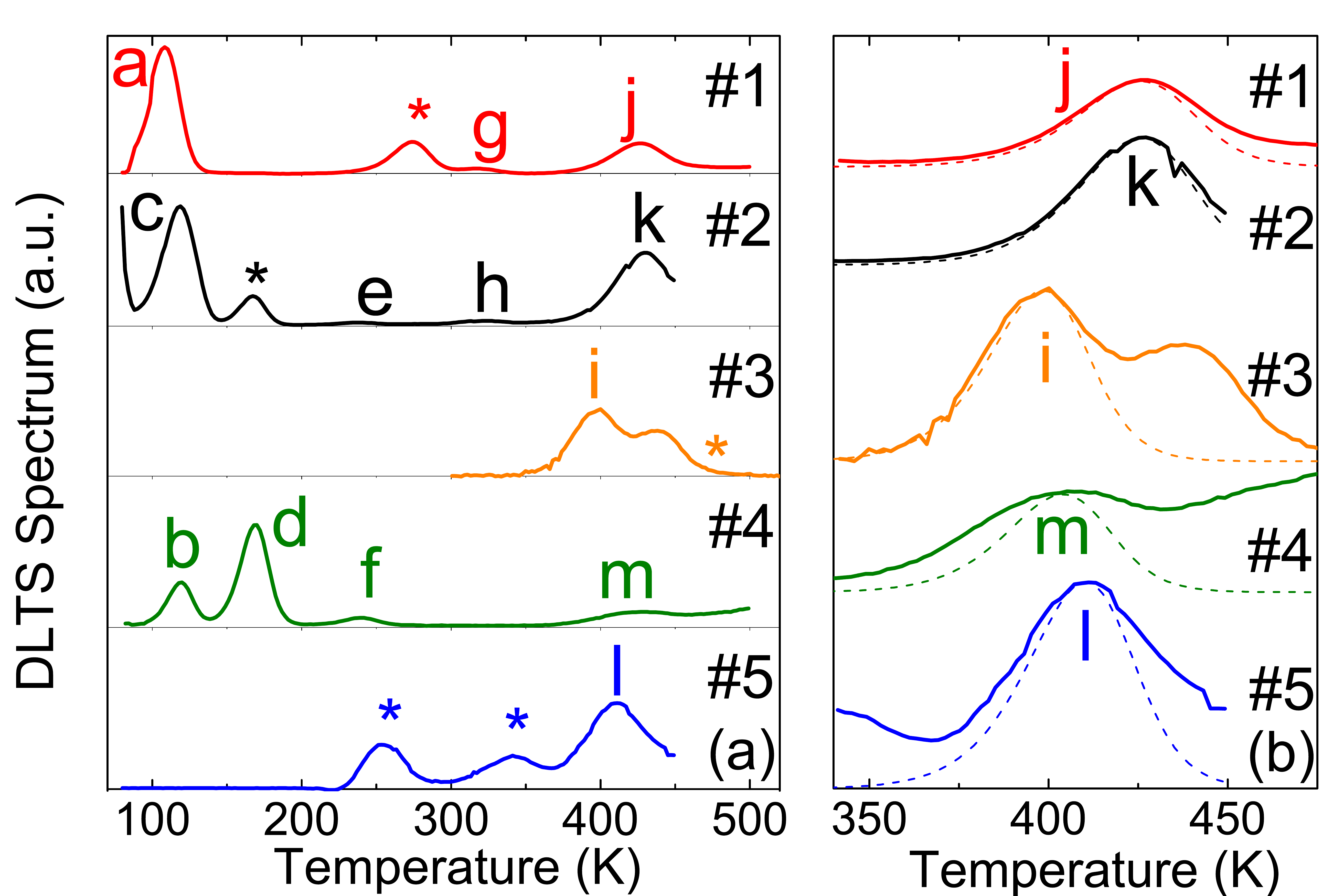}
\caption{ (a) DLTS spectra for each of the five ZnO samples with label of studied peak (peak labeled with star are not studied in this work). Time windows of 100ms  for DLTS spectra of samples $\#1$, $\#2$, 50ms for samples $\#4$, 0.5s for sample $\#5$ and 1s for sample $\#3$ were used on the spectra represented here. (b) E1000 experimental (full line) and simulated (broken line) spectra, the later being calculated with parameters deduced from Arrhenius fit.}
\label{DLTS}
\end{figure}

\begin{figure}
\centering
\includegraphics[width=0.9\linewidth]{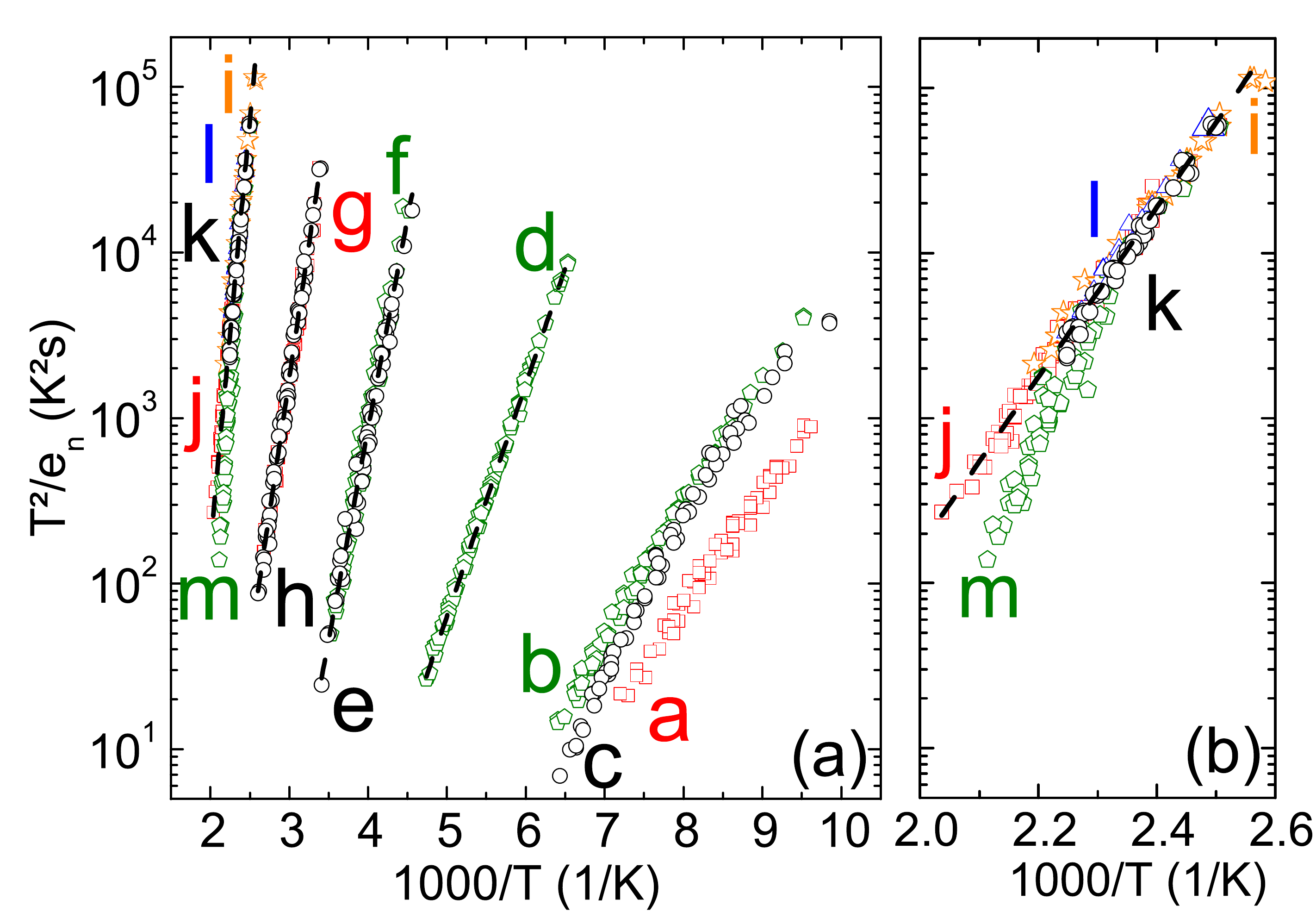}
\caption{(a) Arrhenius diagram of levels measured by Fourier transform deep level transient spectroscopy in five $n-$type ZnO samples. A letter label is attributed to each level. (b) Same as (a) but focused on the E1000 level. Data coming from sample $\#1$ are represented in open squares, $\#2$ in open circles, $\#3$ in open stars, $\#4$ in open pentagons, $\#5$ in open triangles and dash lines correspond to the linear fit of data.}
\label{arrh}
\end{figure}

A total of seventeen electron traps have been detected in the five samples. Typical FT-DLTS spectra are shown on Fig. \ref{DLTS}. The Arrhenius diagrams shown in Fig. \ref{arrh} were obtained by extracting both temperature and emission rates from the maxima detected in DLTS spectra using up to 28 distinct and independent correlation functions yielding back as much Fourier coefficients. Letter labels have been assigned to the thirteen traps found in these five samples(from a to m). Four additional traps are indicated by a star on the figure Fig. \ref{DLTS} and not reported on Fig. \ref{arrh}. These traps will not be discussed in this article. By linear fitting, the activation energy (from the slope) and the apparent capture cross section (from the ordinate at zero abscissa) have been determined using the standard emission rate formula :
\begin{equation}
\label{emissionrate}
e_{n}=\gamma_{n} \sigma_{n} v_{th}N_{C}\exp{\left(\frac{-E_{an}}{k T}\right)}
\end{equation}
where $e_{n}$ is the emission rate, $\gamma_{n}$ the entropy factor assumed to be unity in this section, $\sigma_{n}$  the capture cross section, $v_{th}$ the thermal velocity of electrons, $N_{C}$ the effective density of states in the conduction band, $E_{an}$ the activation energy, $k$ the Boltzmann constant and $T$ the temperature. 

Data falling on the same lines in the Arrhenius diagram of Fig.  \ref{arrh} can be grouped into three ensembles of electron traps ((e,f), (g,h) and (i,j,k,l,m)) labelled  E$_X$, where $X$ is their activation energy in meV. Each of E500, E640 and E1000 is related to a trap with the same physical origin and common to several samples.

The E280 trap is only observed in sample $\#4$ and commonly labelled E3 as reported in literature \cite{Chicot2011}. The electronic  properties (activation energy ($E_{an}$) and apparent capture cross section ($ \sigma_{n}$)) of the four electron traps E280, E500, E640 and E1000 are summarized in Table \ref{TablecaplvlZnO}. A unique fit has been done for each one using the data from the different samples. It must be noticed that the numerous experimental data due to $i)$ the Deep Level Transient Fourier Spectroscopy technique, and to $ii)$ the number of samples, involve very weak error bars in the quantities extracted from the fit, irrespective of systematic errors discussed further.  Three traps (a, b and c) with activation energies from 135 meV to 171 meV and rather low capture cross sections (from $5.3 \times 10^{-18}$ cm$^{2}$ to $4.0 \times 10^{-17}$ cm$^{2}$) has been observed in sample $\#1$, $\#4$ and $\#2$. Some works \cite{auret_2001,auret_2002,auret_2004,monakhov_2009,mtangi_2009} mentioned levels with such low energy but larger capture cross sections except in \cite{scheffler_2011,ye_2011}. The trap E500 observed in sample $\#2$ and $\#4$ is often reported in literature \cite{fang_2008,dong_2010,Quemener,schifano_2009,vines_2010} and commonly named E4 even if its attribution is still unclear. The trap E640 has been observed only in samples $\#1$ and $\#2$, which are two CVT grown samples. This correlation with the preparation method probably means that this trap is linked to a specific impurity of the CVT process. In literature \cite{scheffler_2011, fang_2009}, the rare possible occurrences of this trap are also reported in CVT samples.

The deepest E1000 electron trap is systematically observed in the five samples. The activation energy and apparent capture cross section determined from the Arrhenius diagram have been used to simulate the FT-DLTS spectra shown in figure \ref{DLTS} (b), both obtained from the first real Fourier coefficient \cite{weiss_deep_1988}. For the extraction of $E_{an}$ and $ \sigma_n$ values of the E1000 trap, data coming from the sample $\#4$ were not taken into account. As it is shown on the figure \ref{DLTS}, the peak of the E1000 level of sample $\#4$ is rather broad and probably contains the contribution of other levels, so that the relationship between the emission rate and the peak position of the FT-DLTS spectra becomes inaccurate.

For the four other samples, the good agreement between the experimental spectrum and the simulated one, $i)$ confirms the confidence given by the error bars and $ii)$ indicates that the E1000 trap is a simple point defect in contrast to extended defects which generally result in more broadened spectra with respect to the Fourier transform of a purely exponential transient. The existence of this trap in the five samples, whatever the growth technique and set-up, suggests a native defect (like interstitial, vacancy and related complexes) rather than a foreign impurity, which would have hardly to be common to all these samples, as the origin of this level. The huge capture cross section ($\sigma_n =1.6\pm 0.4 \times 10^{-13}$ cm$^{2}$) deduced in this work clearly indicates that the E1000 trap is attractive for electrons and related to a positive charged centre, and therefore to a donor level. The intensity of the peak is higher in the implanted sample $\#5$ than in other samples (trap concentrations for sample $\#1$, $\#2$, $\#3$, $\#4$ are respectively equal to $2.6 \times 10^{14}$ cm$^{-3}$, $4.2 \times 10^{14}$ cm$^{-3}$, $6.4 \times 10^{14}$ cm$^{-3}$, $1.2 \times 10^{14}$ cm$^{-3}$ versus $3.65 \times 10^{15}$ cm$^{-3}$ for sample $\#5$). In samples $\#1$, $\#2$, $\#3$ and $\#4$, a concentration of native defects like oxygen vacancy V$_\textrm{O}$ close to a few $10^{14}$ cm$^{-3}$ may be the equilibrium one after growth and annealing, whereas it is well known that the implantation process is able to create more vacancies, which cannot be completely annealed out. When passing from sample $\#4$ to $\#5$ (same samples but $\#5$ has been implanted), the concentration of E1000 was multiplied by a factor of more than 30, and the corresponding peak clearly emerges out of the corresponding broad band in sample $\#4$ as shown in Fig. \ref{DLTS} (b). Since the oxygen vacancy V$_\textrm{O}$ is the only deep donor in this ionization energy range with a negative U behavior and a (2+/0) charge transition \cite{Janotti2005, Erhart, Janotti2007, Oba, Lany1, Lany2}, its properties are discussed more deeply in the next section.

\subsection{Electron emission from the oxygen vacancy V$_\textrm{O}$}
Among native point defects, the oxygen vacancy V$_\textrm{O}$ is the only centre which shows thermodynamic transition levels calculated by $ab$ $initio$ methods in the upper half of the band gap in most studies \cite{Erhart, Janotti2005, Oba, Janotti2007}. An other team of theorists \cite{Lany1, Lany2} found these transition levels in the lower half of the band gap, although all these authors agree both about the a$_1$ symmetry of the V$_\textrm{O}$ states, mainly coming from the 4s dangling bonds of the four Zn neighbors which have essentially a conduction band character, and the double donor nature of V$_\textrm{O}$ with a negative correlation energy, making V$_\textrm{O}^{2+}$ and V$_\textrm{O}^{0}$ the only stable states. The half of the two electrons transition energy $(E_C-E_{T}^{2,0})$ (where the upper index holds for the numbers of trapped electrons before and after the transition) is close to 1.2 eV in the studies published by the former authors. When the Fermi level is between the one electron transition energies $(E_C-E_{T}^{1,0})$ and $(E_C-E_{T}^{2,1})$, the formation energy of V$_\textrm{O}^{1+}$ is always higher than those of V$_\textrm{O}^{0}$ and V$_\textrm{O}^{2+}$, thus making V$_\textrm{O}^{1+}$ unstable. In FT-DLTS, after the capture process resulting from the pulse voltage which makes the trap neutral, since the electron involved in the first ionization is bound more strongly than the second one, the second electron emission follows immediately the first one at a given temperature, resulting in a single peak in the DLTS signature with an amplitude multiplied by two \cite{Watkins}. Therefore, the single thermal activation energy E$_{an}$ measured in FT-DLTS, which is due to a single electron emission from V$_\textrm{O}^{0}$ may be determined by $(E_C-E_{T}^{2,1})$, which is larger than $(E_C-E_{T}^{1,0})$, the remaining electron being consequently emitted much faster. The situation was different for the initial and final charge states of defects with negative correlation energy previously known in other semiconductors, because they were either amphoteric in Si \cite{Watkins} or double acceptors lying close to the conduction band in 4H-SiC \cite{Son}, or double donors lying close to the valence band in Si \cite{Watkins}. Hence, a detailed analysis of the capture cross section derivation and theoretical calculations already published has to be addressed because it can be helpful to either validate or discard the assignment of the present E1000 level to V$_\textrm{O}$.

\subsubsection{Electron emission related to the V$_{\textrm{O}}^{0/1+}$ and V$_{\textrm{O}}^{1+/2+}$ transitions}

Taking both capture and emission kinetics and semiconductor statistics into account, the emission rates of the two transitions at thermodynamic equilibrium can be expressed respectively as 
\begin{equation}
\label{en2}
e_{n2}=\gamma_{1}\sigma_{n1}v_{thn}N_{C}\exp(-\Delta H_{T}^{2,1}/kT)
\end{equation} 
for the V$_{\textrm{O}}^{0/1+}$ transition and 
\begin{equation}
\label{en1}
e_{n1}=\gamma_{0}\sigma_{n0}v_{thn}N_{C}\exp(-\Delta H_{T}^{1,0}/kT) 
\end{equation}
for the V$_{\textrm{O}}^{1+/2+}$ transition, 
where $\sigma_{n,i}$ is the capture cross section, $\gamma_{i}$ the entropy factor discussed in the following, $\Delta H_{T}^{i+1,i}$ the enthalpy of the transition, $i$ the number of trapped electrons, with $i=0$ in the case of V$_{\textrm{O}}^{2+}$. The entropy factor $\gamma_{i}$ has two contributions: the degeneracy factor and the vibrational entropy. The neutral state V$_{\textrm{O}}^{0}$ is obtained if four electrons lie in the four a$_1$ states, with a total degeneracy of $m=8$. Consequently, the configuration parts of the degeneracy factor, equal to the ratio of the number of possible combinations C$_{m}^{i+3}$/C$_{m}^{i+2}=\frac{m-i-2}{i+3}$ are respectively 5/4 for $\sigma_{n1}$ and 2 for $\sigma_{n0}$. The other part $\exp(\Delta S_{i,vibr}/k)$ of the entropy factor is due to vibrational entropy which reaches its maximum for band to band transitions and determines the temperature dependance $dE_{G}/dT$ of the band gap energy $E_{G}$ \cite{Thurmond, ODonnell}. From measurements of Rai and co-workers \cite{Rai}, $dE_{G}/dT$ is close to 0.1 meV/K in the range 300-450 K, a value which induces upper limits of 1.25 for $\Delta S_{E_{G},vibr}/k$ and 3.5 for $\exp(\Delta S_{E_{G},vibr}/k)$. But the effective entropy change in the transition $\Delta S_{i,vibr}$ is much smaller because the levels are expected to follow the conduction band edge from which the states of V$_\textrm{O}$ originate and the transition energy is close to only one third of the band gap energy $E_G$. Therefore, the total entropy factors $\gamma_{0}$ and $\gamma_{1}$ would be very close to one, as previously assumed.

\subsubsection{Capture cross section of the V$_{\textrm{O}}^{0/1+}$ and V$_{\textrm{O}}^{1+/2+}$ transitions}

In equations \ref{en2} and \ref{en1}, the capture cross sections $\sigma_{n1}$ and $\sigma_{n0}$ are dependent on the microscopic properties of the defect or impurity, and temperature in the general case. Extensive theoretical calculations of the multi-phonons mediated transition probability per unit time and capture cross section of deep levels have been performed within the framework of the Born-Oppenheimer approximation from the seventies to the nineties \cite{Passler1, Ridley, Henry, Passler2, Bourgoin, Brousseau, Goguenheim}. These quantities are dependent on the average phonon energy $\hbar \omega_q $ in a relative way because $\hbar \omega_q $ has to be compared to the thermal energy $kT$, to the ionization enthalpy $\Delta H_{T}^{i+1,i}=p_i \hbar \omega_q $ thus defining the number of phonons $p_i$ necessary for energy conservation and to the Condon shift $S_i \hbar \omega_q $ where the Huang and Rhys factor $S_i$ scales the coupling between the phonon modes and the one electron states. The picture which emerged from these calculations allows to write the capture cross section as:
\begin{equation}
\sigma_{ni}= \sigma_{ \infty,i}~F( \hbar \omega_q /kT, p_i, S_i)~C_Z
\end{equation}
 where $ \sigma_{ \infty,i}$ is the capture cross section limit for infinite temperature of a neutral center which directly depends on the matrix element of the non-adiabatic hamiltonian between initial and final states, $F( \hbar \omega_q /kT, p_i, S_i)$ the line shape function of the optical spectrum for a zero photon energy and $C_Z$ the averaged Sommerfeld factor, which takes into account the deformation of the wave functions induced by the Coulomb potential \cite{Passler1, Brousseau}. From the analytic expression given in reference \cite{Passler2}, the product $ F(\hbar \omega_q /kT, p_i, S_i)~C_Z$ can be calculated and a thermal activation of the cross section can be inferred. For the oxygen vacancy in ZnO, the Huang and Rhys parameter can be assessed from the configuration diagram given in the Fig. 3 of ref. \cite{Janotti2005}: close to $p_1/2$ in the transition V$_{\textrm{O}}^{1+/0}$ which involves the $\sigma_{n1}$ capture cross section measured in DLTS experiments and close to $p_0$ in the transition V$_{\textrm{O}}^{2+/1+}$. Generally speaking, the prefactor $\sigma_{\infty,i}$ is more difficult to assess because it is proportional to the matrix elements of the perturbative hamiltonian. It has been estimated in the range $10^{-15}-10^{-14}$ cm$^{2}$ by Henry and Lang \cite{Henry} for most impurities and is taken to be 10$^{-14}$ cm$^{2}$ by P\"{a}ssler \cite{Passler1}. But as shown by Ridley \cite{Ridley}, it is both proportional to $S_{i}^{2}$ and then increases with the electron-phonon coupling, which is rather high in V$_\textrm{O}$ since $S_i$ are of same magnitude as $p_i$ or $p_i/2$, and to the matrix element of the perturbative hamiltonian calculated by an integral over spatial coordinates of the wave functions of the bound electron and delocalized one. In the case of a vacancy, the wave function of the bound electron spreads over a much larger distance than for an impurity center because it is localized in the dangling bonds of the neighbouring atoms. This fact justifies that $\sigma_{ \infty,1}$ must amount to about $2 \times 10^{-13}$ cm$^{2}$ in order to fit the experimental value. The capture cross section $\sigma_{n0}$ involved in the transition V$_{\textrm{O}}^{2+/1+}$ is expected to be even greater because the Huang and Rhys parameter $S_0$ and $Z$=2 are both higher than in the V$_{\textrm{O}}^{1+/0}$ transition.

\subsection{Discussion}

The emission rate inequality $e_{n2} \ll e_{n1}$ is confirmed both because $\sigma_{n1} < \sigma_{n0}$ and  $\Delta H_{T}^{2,1} > \Delta H_{T}^{1,0}$, implying that only the slower ($e_{n2}$) emission events are detected in DLTS at all temperatures. This means that the electronic transition of the oxygen vacancy measured by DLTS is only characteristic to the V$_{\textrm{O}}^{0/1+}$ transition. Also, the capture cross section of the transition must correspond to a single positively charged center and the activation energy $E_{an}$  deduced from the Arrhenius diagram must be compared to the $(0/+)$ transition calculated by \textit{ab initio} methods. 

It must be noticed that the capture cross section which is deduced from an Arrhenius diagram, is neither the effective one in the measurement temperature range nor the theoretical value $\sigma_{ \infty,i}$ at infinite temperature but an intermediate value obtained at the intercept of the tangent to the curve with the vertical axis located at infinite temperature. Consequently, in the case of  $\sigma_{ \infty,1}$, the effective capture cross section in the temperature range of measurements is smaller but still in the range of the value given by the Arrhenius diagram due to the thermal activation of the $ F(\hbar \omega_q /kT, p_i, S_i)~C_Z$ factor. Despite such a lowering, the real capture cross section cannot be measured directly because, taking into account the net doping concentrations given in table \ref{tab:descriptifEch}, the typical capture kinetic amounts to only  some picoseconds (too short for measurement). Anyway, the order of magnitude of the capture cross section ($\gtrsim10^{-13}$ cm$^{2}$) of the trap measured in this work is in good agreement with a positively charged center (attractive for electron) like the V$_{\textrm{O}}^{+}$ which is the only native defect being related to an electronic state within the band gap with an attractive character for electrons in ZnO. 

The measured activation energy are weaker than those calculated in previously quoted theoretical studies \cite{Janotti2005, Erhart, Janotti2007, Oba} by some tenths of eV for the V$_{\textrm{O}}^{1+/0}$ transition \cite{notice} . But both because the systematic presence of the E1000 trap implies an assignment to a native defect rather than an impurity and the oxygen vacancy is the only native defect which is an attractive centre for electrons, an 1 eV value assigned to the enthalpy of the V$_{\textrm{O}}^{1+/0}$ transition of the oxygen vacancy is most probable. Moreover, the discrepancy between experimental and most of theoretical transition enthalpies \cite{Janotti2005, Erhart, Janotti2007, Oba} is noticeably smaller in comparison with Hofmann \textit{et al.} proposal \cite{Hofmann} which assigned the electron trap at $E_C-530$ meV to the oxygen vacancy. It must be noticed that the assignment of the level at $E_C-530$ meV to the oxygen vacancy is discarded by two important results reported in this work: $i)$ this level is found here (E500) in only two of the five samples studied in this work which is inconsistent with a native point defect like the oxygen vacancy, $ii)$ this level is found in sample $\#4$ (not implanted) and not in sample $\#5$ (same than $\#4$ but implanted in the case of $\#5$), in contradiction with the assignment to an oxygen vacancy created by implantation process.

The energy level reported by Quemener  \textit{et al.} \cite{Quemener} for the E5 trap is in good agreement with our results and seems to be also the V$_\textrm{O}$, except that the capture cross section is slightly lower. Unfortunately, the Arrhenius diagram has not been shown, preventing the detailed comparison with the five Arrhenius diagrams reported in the present work. Future works will be needed to confirm the assignment of the V$_{\textrm{O}}^{1+/0}$ to the E1000 trap. Indeed, the possibility that the E1000 trap is related to a complex between an impurity (present in ZnO whatever the growth method, most probably H) and a native defect (created by implantation for example) cannot be completely discarded. However, the good agreement between the experimental data reported here for all the five different samples which permitted DLTS measurements up to 450 K and the \textit{ab initio} calculations makes the oxygen vacancy the best candidate for the E1000 trap.

\section{Conclusions}
Traps have been detected and their parameters extracted from FT-DLTS data in five different $n-$ZnO samples. A salient feature of this work consisted to show that only the deepest level ever detected by an electrical method, labelled E1000, with a ionization energy of 1 eV for electrons and a very probable donor character due to its very large capture cross section, turns out to be present in all the five samples, in contrast to other deep levels. From a detailed analysis of the electronic properties of the multi-phonons mediated transitions taking place in the oxygen vacancy and comparison with our experimental results, we can conclude that the capture cross section and ionization energy deduced from the experimental Arrhenius diagram are compatible with those estimated from theoretical considerations pertaining to the oxygen vacancy in ZnO, which is recognized as a double donor with a negative correlation energy. 

\acknowledgements
The authors are grateful to Michel Lannoo for valuable discussions.

\newpage

nn,
\newpage

\begin{table*}[!htbp]
  \centering 
\begin{tabular}{p{1,4cm}p{5cm}p{3cm}p{3cm}}
\hline
\hline 
Sample &  Growth method/Origin&Remark  &  $N_{d}-N_{a}$ ($cm^{-3}$) \\
\hline
$\#1$    & CVT on ZnO/Leti-CEA     & homoepitaxial 	&3.0$\times10^{+15}$\\
$\#2$    & CVT on saphir/Leti-CEA  & heteroepitaxial 	&1.6$\times10^{+16}$\\
$\#3$    & HT/Crystec Inc.                & no 			&1.2$\times10^{+16}$\\   
$\#4$    & HT/Tokyo Denpa Inc.       & no  			& 3.0$\times10^{+16}$\\
$\#5$    & HT/Tokyo Denpa Inc.       & implanted		&2.3$\times10^{+16}$\\
 \hline
 \hline 
\end{tabular}
  \caption{Description of the five samples investigated in the work (growth method, origin of the sample, additional remarks and effective doping $N_{d}-N_{a}$ evaluated from C(V) measurement). }\label{tab:descriptifEch}
\end{table*}

\begin{table*}
\begin{tabular}{p{1,5cm}p{2cm}p{3cm}p{3cm}c}
\hline 
\hline 
Trap & Label & $E_{an}$ (eV) 	&$ \sigma_n$ (cm$^{2}$) & Samples\\
\hline
E280& d                &$0.278$ & 	$1.8\times10^{-16}$	&$\#4$ \\
E500& e ,f             &$0.505$ &	$2.5\times10^{-14}$	&$\#2$, $\#4$ \\
E640& g, h            &$0.644$ &	 $4.6\times10^{-15}$	&$\#1$, $\#2$\\
E1000& i, j, k, l, m &$1.018$ & 	$1.6\times10^{-13}$	&$\#3$, $\#1$, $\#2$, $\#5$, $\#4$\\
\hline
\hline 
\end{tabular}
\caption{Activation energy ($E_{an}$) and capture cross section ($ \sigma_n$) of electron traps detected in the five different ZnO samples investigated in this work. The labels correspond to the ones of the Arrhenius diagram and DLTS spectra.}\label{TablecaplvlZnO}
\end{table*}


\begin{thebibliography}{9}
\bibitem{notice} The numerical simulations showed that the thermal activation of the cross section, involved in the $ F(\hbar \omega_q /kT, p_i, S_i)~C_Z$ factor, can be neglected in comparison with the activation energy $E_{an}$ measured in this work considering phonon energy at 40 meV.
\end{thebibliography}
\end{document}